\documentclass[sigconf, nonacm]{acmart}
\usepackage{array}
\usepackage{xcolor}
\usepackage[normalem]{ulem}
\definecolor{revmcol}{rgb}{0.72,0.00,0.55}

\newcommand{\tabnote}[1]{%
  \smallskip\par
  \begingroup
  \leftskip=0pt\rightskip=0pt\parfillskip=0pt plus 1fil\relax
  \noindent\footnotesize #1\par
  \endgroup}

\AtBeginDocument{%
  }

\setcopyright{none}
\copyrightyear{2026}
\acmYear{2026}
\acmDOI{}
\acmISBN{}
\renewcommand\footnotetextcopyrightpermission[1]{}
\author{Jacky Chen}
\affiliation{%
  \institution{University of Toronto}
  \city{Toronto}
  \country{Canada}
}
\email{jackyjc.chen@rotman.utoronto.ca}

\author{Ing-Haw Cheng}
\affiliation{%
  \institution{University of Toronto}
  \city{Toronto}
  \country{Canada}
}
\email{inghaw.cheng@rotman.utoronto.ca}

\author{John Hull}
\affiliation{%
  \institution{University of Toronto}
  \city{Toronto}
  \country{Canada}
}
\email{john.hull@rotman.utoronto.ca}

\author{Landis Li}
\affiliation{%
  \institution{University of Toronto}
  \city{Toronto}
  \country{Canada}
}
\email{landisw.li@utoronto.ca}

\author{Zissis Poulos}
\affiliation{%
  \institution{York University}
  \city{Toronto}
  \country{Canada}
}
\email{zpoulos@yorku.ca}

\author{Justin Shi}
\affiliation{%
  \institution{University of Toronto}
  \city{Toronto}
  \country{Canada}
}
\email{justinzy.shi@utoronto.ca}

\author{Andreas Veneris}
\affiliation{%
  \institution{University of Toronto}
  \city{Toronto}
  \country{Canada}
}
\email{veneris@eecg.utoronto.edu}

\author{Zeyu Wang}
\affiliation{%
  \institution{University of Toronto}
  \city{Toronto}
  \country{Canada}
}
\email{colinzeyu.wang@utoronto.ca}

\renewcommand{\shortauthors}{J. Chen, I.-H. Cheng, J. Hull, L. Li, Z. Poulos, J. Shi, A. Veneris, and Z. Wang}
\begin{document}

\title{Risk-Sensitive Reinforcement Learning for an Options Market Maker: An Empirical Study with Real S\&P 500 Options}

\begin{abstract}
Options market makers absorb client order flow and must hedge the resulting inventory risk under trading frictions. We study whether an option hedging reinforcement learning (RL) agent trained on historical market data outperforms three alternatives in tail-risk protection: (i) traditional delta hedging, (ii) simple hedge-ratio rules, and (iii) the same agent trained in simulated markets. Most RL hedging agents are trained in model-generated markets that do not capture the crashes and volatility regimes present in real data. We introduce an empirical replay environment using S\&P~500 option data from 1996 to 2017 and compare against a matched geometric Brownian motion (GBM) environment, a stochastic-volatility (SABR) simulator, and a hybrid that values historical paths with Black--Scholes, all sharing the same portfolio mechanics, stochastic arrivals, costs, and test period. We train a distributional Twin Delayed Deep Deterministic Policy Gradient (TD3) agent and evaluate it out of sample over 2018--2023. The agent trained on empirical data reduces Conditional Value-at-Risk (CVaR) tail loss by $3.1\%$ compared to delta hedging and outperforms every hedge-ratio rule chosen during the training period and every alternative training environment. The reduction reaches $8.9\%$ in volatile and falling markets, but reverses to a cost of $7$--$10\%$ in rising markets. Empirical training induces a state-dependent short net delta: the policy under-hedges negative-delta books, most strongly after prices have fallen. This tilt disappears with GBM training and is weaker with SABR or hybrid training. These findings suggest that exposure to historical market dynamics can materially change a tail-risk hedging policy, although the evidence rests on a single test path and the edge is regime-conditional.

\end{abstract}


\maketitle

\section{Introduction}

For options market makers, delta hedging is the standard method for managing the directional risk of clients' option portfolios. When taking on client orders, market makers build option positions and must manage the associated risks~\cite{ho1981optimal,garleanu2009demand}, which is a central challenge in market microstructure. Traders offset the delta of the option portfolio by trading the underlying asset, but textbook rules are derived under the premise of continuous and frictionless trading. In reality, hedging adjustments are discrete, transaction costs are high, and option portfolios change with the arrival of client orders, which cluster on large-move days. Therefore, maintaining precise delta neutrality on each rebalancing day may not be the optimal balance between transaction costs and portfolio risk.

Reinforcement learning (RL) provides an alternative to delta hedging: an RL agent learns a state-dependent hedging ratio from partial information instead of fixing the ratio to 1. Previous research has shown that RL can outperform delta hedging under transaction costs and has been applied to market making~\cite{spooner2018market}, but much of this evidence comes from agents trained and evaluated in a model-generated environment. For example, \citet{cao2021deep} study a call option in geometric Brownian motion (GBM) and SABR environments. A constant-volatility model such as GBM provides a controlled setting but cannot reproduce the crashes and volatility clustering seen in historical data. This raises an important question for AI-driven market making: can an agent trained on historical market data, rather than simulated data, manage inventory risk better than conventional delta hedging?

We break this into three main questions: (1) Can an RL agent trained on historical option data outperform conventional delta hedging when both are evaluated on unseen market data? (2) Does the agent also outperform hedge-ratio strategies calibrated on the same training data? (3) When tested over the same historical period, does an agent trained on historical data outperform a same-setup agent trained on a GBM simulator, and is this performance difference coming from the price paths or from option valuation?

To study these questions, we build an empirical environment using the S\&P~500 (SPX) index and options data that replay the historical path. We train RL agents in this environment with data from 1996 to 2017 and evaluate their performance outside the sample with data from 2018 to 2023. We compare this environment with a GBM simulation environment, and with a hybrid environment that is identical to the empirical one except that options are valued with Black--Scholes, using the same portfolio mechanism, stochastic client option arrivals, transaction costs, state and action spaces, and reward. This design provides three comparisons: empirically trained RL vs. delta hedging, empirically trained RL vs. simple hedge-ratio rules, and empirically trained RL vs. GBM-, SABR- and hybrid-trained RL. The last one addresses a broader question for AI in market microstructure: how much a trading agent's behavior depends on whether it learns in a simulator or on replayed market data.

Our primary RL learner is the distributional Twin Delayed Deep Deterministic Policy Gradient (TD3), where we improve on TD3 by replacing the scalar critics with quantile critics so the agent can directly optimize the tails of the value distribution. TD3's twin critics and delayed actor and target-network updates are designed to reduce optimistic value estimates, which can be harmful when tail observations are scarce. These conservative mechanisms are our reason for building on TD3 rather than a plain distributional actor--critic such as Distributed Distributional Deep Deterministic Policy Gradient (D4PG), which has been used in prior option hedging work~\cite{barth2018distributed,cao2023gamma} but lacks the twin-critic minimum and delayed updates. In our experiments, we keep the learner fixed and vary the training environment to isolate the impact of the training environment. A detailed analysis of the specific contributions of TD3's individual components is left for future research.

During the 2018--2023 test period, the TD3 agent trained on historical data achieves a $95\%$ Conditional Value-at-Risk (CVaR$_{95}$) of $-77.7$, averaged over five independent training runs, while delta hedging achieves a CVaR$_{95}$ of $-80.2$, which corresponds to a reduction of approximately $3.1\%$ in tail loss. The benefit is much larger in volatile markets that are in decline, where tail loss is reduced by $8.9\%$. The tradeoff is that the strategy reverses to a cost of $7$--$10\%$ in rising markets. The agent also outperforms six hedge-ratio rules chosen on the training data. The learned policy maintains a state-dependent short net delta: it under-hedges portfolios with negative delta, most strongly after the index has fallen. This creates a general tilt toward protection against market declines.

Comparative analysis across environments further reveals that agents trained on historical data achieve the largest tail-risk improvement of all training environments, with these gains concentrated in high-volatility falling markets. The GBM-trained counterpart does not beat delta hedging, while an agent trained in a stochastic-volatility (SABR) simulator and a hybrid agent trained on the same historical paths but with Black--Scholes option values each recover only part of the gain; in our runs the largest gain comes from training on both historical price paths and market option quotes.

\section{Related Literature}
\paragraph{Delta hedging and RL under market frictions.}

The Black-Scholes-Merton (BSM) model provides a classic benchmark for the pricing of European options and delta hedging under idealized assumptions \cite{black1973pricing,merton1973theory}. In this benchmark model, traders offset the first-order exposure of the option book by trading the underlying asset. However, trades in practice are made at discrete times with substantial transaction costs and changing client books. As a result, maintaining the exact delta neutrality may not be the best decision over multiple periods \cite{leland1985option}. Market-microstructure models treat this inventory risk as the central cost of liquidity provision~\cite{ho1981optimal,avellaneda2008high}, and option dealers' hedging demand affects option prices~\cite{garleanu2009demand}.

Recent work has explored whether RL can improve derivatives hedging under realistic market frictions. Buehler et al.\ use deep RL to hedge under market frictions such as transaction costs, liquidity constraints or risk limits \cite{buehler2019deep}. Kolm and Ritter use RL to study how a hedging portfolio should be adjusted when trades are discrete and transaction costs are nonlinear \cite{kolm2019dynamic}. Cao et al. show that RL can improve a given hedging objective over delta hedging with transaction costs, and they highlight the accounting profit-and-loss (PnL) reward in the option book, which is valued at market value at each step \cite{cao2021deep}. Subsequent work extends these ideas to gamma and vega hedging, barrier options, and other derivative combinations \cite{cao2023gamma,chen2024barrier}, and RL market makers learn quoting and inventory policies~\cite{spooner2018market}. These works inspire us to adopt delta hedging as the benchmark and RL to learn the hedge ratio, but do not clarify how much the training data source affects the tail risk at the portfolio level.

\paragraph{Model-generated vs. empirical training environments.}

Most RL hedging works train and evaluate agents in model-generated markets, such as GBM, Heston, or other calibrated stochastic processes \cite{cao2021deep}. Their regime structure is determined by the model rather than by the realized market history. Mikkil\"a and Kanniainen demonstrate the importance of empirical option prices for reinforcement learning hedging~\cite{mikkila2023empirical}, and Zernikov examines what such agents learn~\cite{zernikov2026deep}; both study individual options without client flow or a distributional tail-risk objective. Our work addresses this training environment problem.

\paragraph{Distributional RL and tail-risk estimation.}

Tail-risk hedging cannot be well learned from a critic that predicts only the expected return. The distributional RL instead estimates the full distribution of possible returns \cite{bellemare2017distributional}. Quantile-regression methods represent this distribution using a set of quantiles, which makes it possible to read off measures such as Value-at-Risk (VaR) and CVaR \cite{dabney2018distributional}. In our setting, the reward at each step is still the marked-to-market PnL of the hedging book, but the actor chooses its policy based on the CVaR calculated from the critic's estimated return distribution.

The challenge of distributional RL is to estimate the tail of this distribution, since stress events are rare in historical data. The Deep Deterministic Policy Gradient (DDPG) is a deterministic actor-critic RL algorithm designed for continuous actions \cite{lillicrap2016continuous}. TD3 extends DDPG with twin critics, target-policy smoothing, and less frequent actor updates to limit overestimation \cite{fujimoto2018addressing}. D4PG is also an extension of DDPG that uses a distributional critic and has previously been applied to option hedging \cite{barth2018distributed,cao2023gamma}.

\section{Environments}

\subsection{Common Hedging Task}
\label{subsec:common}

The environment assumes a trading desk that holds a client's option portfolio and hedges it by trading the underlying asset. The initial option portfolio remains fixed across episodes: three near-the-money S\&P~500 options, consisting of two short calls and one long call. New client options arrive during the episode through separate Poisson processes for calls and puts, following the arrival rule described below. After each arrival or expiration, the book value and net Greeks are re-aggregated before the next hedging decision.

Each episode simulates 40 trading days with daily rebalancing. At each step, the agent adjusts its stock position to manage the delta exposure of the portfolio. The market then moves forward by one trading day, and the reward is the resulting change in the value of the hedged portfolio, combining the option book and the stock hedge, after trading costs. While an option is live, it is marked to market from market mid-quotes in the empirical environment and analytically in the other training environments. On the first trading day on or after its expiration date, it is settled at intrinsic value and removed from the book; options still live at the 40-day horizon are marked to market, with no terminal liquidation cost. New client options arrive at the mid-quote with no inception PnL. Each stock transaction incurs a percentage transaction cost of 3 basis points (bps), which represents bid-ask spreads and execution slippage. We assume no financing cost on the stock hedge (risk-free rate $r_f=0$). To make learning scale-invariant, the underlying price is rebased on a common initial level $S_0=10$ at the beginning of every episode.

Portfolio construction, client arrivals mechanism, action space, rewards, 3~bps transaction costs, and a 40-day trading horizon are identical across all training environments, and each agent is tested using the same historical replay data from 2018--2023 (Table~\ref{tab:training_env_design}). The environments differ in the source of the underlying path and option valuation data. With this setup, we separate the common hedging task from the training-data comparison.

\subsection{Option Flows}

The environment starts with an existing client option portfolio and allows new client options to arrive during the simulation. Each available client option is a call or a put drawn from the historical option chain with a moneyness of $[0.99,1.01]$ and $20$--$60$ days to maturity. Calls and puts are simulated with Poisson arrival processes and are added to the client portfolio when they arrive, each as a long or short position with equal probability. Upon each arrival, the portfolio value and net Greeks are re-aggregated before proceeding to the next step.

The formulation of the option-flow rule is based on observations from Chicago Board Options Exchange (CBOE) trading volume data: option trading activity is more intense on days of high market volatility than on days of market stability. Our model does not replicate more granular volume asymmetries, such as those between call and put options or between upward and downward price movements, as the order arrival process employed remains symmetric with respect to contract type and directional outcome.

We turn this into a two-level Poisson arrival process with the same intensity for calls and puts. We use the high-arrival intensity, $\lambda_{\mathrm{high}}$, when the absolute one-day index return exceeds one daily standard deviation implied by the volatility parameter of the environment ($\sigma/\sqrt{252}$), and the low-arrival intensity, $\lambda_{\mathrm{low}}$, otherwise; for the baseline we set $\lambda_{\mathrm{high}}=0.5$ and $\lambda_{\mathrm{low}}=0.25$. The return that triggers the intensity at step $t$ is the realized return in $t$ (computed from $S_t$ and $S_{t-1}$), so the arrival rule does not use future information. Because the threshold is set by each environment's volatility parameter, it corresponds to a slightly different absolute move across training environments (about $1.01\%$ in the empirical and hybrid environments, $1.20\%$ in the GBM and SABR environments, and $1.32\%$ in the oracle-$\sigma$ GBM); every agent is tested at the empirical threshold, and the functional arrival rule is otherwise identical, so the client-arrival mechanism is held nearly fixed across the environment comparison.

\subsection{Training Environments}

We compare the GBM simulator environment with an empirical replay environment. They differ in two controllable dimensions: the source of the underlying asset price path and the source of the option value and Greeks. Therefore, this comparison identifies the impact of the overall training environment, rather than the isolated effects of historical price paths; the hybrid environment described below separates the option valuation from the rest of the environment.

\begin{table}[t]
\centering
\caption{Training-environment design (shared settings in Section~\ref{subsec:common}).}
\label{tab:training_env_design}
\footnotesize
\setlength{\tabcolsep}{2.0pt}
\renewcommand{\arraystretch}{1.05}
\begin{tabular}{@{}>{\raggedright\arraybackslash}p{0.165\columnwidth}>{\raggedright\arraybackslash}p{0.281\columnwidth}>{\raggedright\arraybackslash}p{0.223\columnwidth}>{\raggedright\arraybackslash}p{0.281\columnwidth}@{}}
\toprule
\textbf{Component} & \textbf{Empirical replay} & \textbf{Hybrid} & \textbf{GBM / SABR} \\
\midrule
Training paths & Historical S\&P~500, 1996--2017 & As empirical & Simulated on the same normalized price scale (SABR: stochastic volatility) \\
Option values and Greeks & SPX mid-quotes with OptionMetrics implied volatility and Greeks & Black--Scholes at a constant volatility of $0.19$ & Black--Scholes at the simulated volatility \\
Client contracts & $20$--$60$-day contracts sampled from the historical SPX chain & As empirical & Near-the-money, $20$--$60$-day maturity \\
\bottomrule
\end{tabular}
\end{table}

\paragraph{GBM simulation environment.}

In the GBM environment, the underlying asset price path is generated by a geometric Brownian motion process with zero drift and constant volatility (\(\sigma=0.19\), the annualized realized volatility of daily log returns over the training period $1996$--$2017$) and an initial price level identical to the normalized initial price of the empirical environment (\(S_0=10\)). The evaluation period $2018$--$2023$ is more volatile (annualized realized volatility $\approx 0.21$), which is part of why GBM-trained agents transfer imperfectly. Since there is no historical option chain in the GBM setting, the option book is priced analytically.

By construction, GBM training data lack empirical characteristics of the stock index market, such as clustering of volatility, jumps, regime changes, and prolonged crisis periods.

\paragraph{Empirical replay environment.}

In the empirical replay environment, the underlying price path is taken directly from the historical S\&P~500 index. Its options market has a long track record of being highly liquid during both periods of market tension and calm. The data set includes the early-2000s dot-com crash, the 2008 global financial crisis, the COVID-19 crash, and the 2022 rate-hike period.

This environment does not simulate index returns. Instead, it reconstructs each episode using a 40-day historical window. Consequently, it preserves the sequence of realized returns while allowing for the sampling of multiple realized paths of option flows from the same historical period.

We divide the historical data set into a training period from 1996 to 2017 and an evaluation period from January 2018 to August 2023 to ensure that no data from the evaluation period are seen during training while including enough stressed market events in both periods. Each training episode begins on a randomly sampled start day in the training period and replays the 40-day trading window that starts on that day.

When constructing client option portfolios, the historical replay environment utilizes market-observed option data from OptionMetrics. The environment does not generate synthetic contracts. Instead, it extracts each new option contract from the available near-the-money contracts in the historical option chain of that day. Contract sampling is volume-weighted, making contracts with higher liquidity more likely to be chosen. The implied volatilities are calibrated by OptionMetrics, and the relevant contract-level Greek data also come from the same records; no recalculation is performed in the replay environment. The book Greeks are the sum of the contract-level Greeks across current positions.

For tail-risk objectives, empirical replay allows the agent to train during actual periods of market stress when poor hedging decisions could result in substantial losses.

\paragraph{Hybrid environment.}
The hybrid environment is identical to the empirical environment, with the same historical index paths, listed contracts drawn from the day's option chain, and client-arrival rule, except that the option book and its Greeks are valued with Black--Scholes at a constant volatility of $0.19$ (the training-period realized volatility) and zero rates. This isolates the effect of valuing the book on market prices during training.

\paragraph{SABR environment.}
To test a richer simulator, we also train in a stochastic-volatility environment SABR~\cite{hagan2002managing} that extends the GBM environment: the volatility follows a lognormal process with volatility of volatility $0.6$, correlation $\rho=-0.7$ with the price (a leverage effect) and initial level $0.19$ in every episode, and options are valued with Black--Scholes at the current volatility.

\section{Methodology}

\subsection{Problem Formulation}

We formulate the hedging problem as a finite-horizon partially observable Markov decision process (POMDP)
\[
    \mathcal{M}=(\mathcal{S},\mathcal{O},\mathcal{A},P,r,\gamma),
\]
where \(\mathcal{S}\) denotes the underlying state space, \(\mathcal{O}\) the observation space,
\(\mathcal{A}\) the action space, \(P(s_{t+1}\mid s_t,a_t)\) the transition kernel induced by the
market path, the arrival of options and the evolution of the portfolio,
\(r\) the one-step
reward function and \(\gamma\) the discount factor. The underlying state \(s_t\in\mathcal{S}\),
which comprises the market path, the complete client book and the position of the carried stock of the desk,
evolves as a Markov process. The agent does not condition on \(s_t\) directly; instead, it acts on a
compact observation \(o_t\in\mathcal{O}\) defined below. At each time step \(t\), the agent
receives \(o_t\), selects a stock-hedging action \(a_t\in\mathcal{A}\), and receives a
mark-to-market portfolio reward \(r_t=r(s_t,a_t,s_{t+1})\) after the market advances by one
trading day. Each episode lasts \(T=40\) trading days and the portfolio is rebalanced daily.
We set
\(\gamma=1\) because the horizon is short and the rewards are measured in the daily marked-to-market PnL.

Because \(o_t\) discards part of \(s_t\), it is not a sufficient
statistic for the underlying state. We therefore restrict our attention to {memoryless}
policies \(\pi:\mathcal{O}\to\mathcal{A}\) that map the current observation to an
action and optimize within this class rather than over history-dependent policies. All
learning quantities in Section~\ref{subsec:dtd3} are consequently written as functions of \(o\)
rather than \(s\).

At time \(t\) the agent receives a compact normalized summary of the market and the condition of the portfolio. In our empirical environment setup, this is the two-dimensional
vector
\[
    o_t = \big(\tilde{S}_t,\ \tilde{\delta}^{\mathrm{book}}_t\big),
\]
the normalized underlying price \(\tilde{S}_t\) and the net delta exposure \(\tilde{\delta}^{\mathrm{book}}_t\) of the client option book, each Min--Max normalized to a fixed range. The
observation excludes the desk's carried stock position \(h_{t-1}\) and is computed {after} expiring options are settled and new client options arrive. As a result, the action is chosen using the exposure of the post-arrival portfolio.

These two variables are the main input for the actual decision that the agent 
makes. The action rescales the delta-neutral target \(\Delta^{\mathrm{target}}_t=-\delta^{\mathrm{book}}_t\),
so the book delta determines the neutral hedge. The underlying price provides information about the options' moneyness and, therefore, about how quickly that delta will move. We deliberately keep the
observation small because it reduces the risk of overfitting to the limited number of independent stress episodes in the historical training data. A compact observation also makes the learned policy easier to interpret, which we exploit
in Section~\ref{subsec:tilt} when we characterize behavior over the book-delta and price deciles.

The action \(a_t\in\mathcal{A}\) is a scalar hedge ratio constrained to
\[
    \mathcal{A}=[-0.2,2.0].
\]
Let \(\Delta_t^{\mathrm{target}}\) denote the position of the stock that would make the delta of the client's option book neutral at time \(t\); if \(\delta_t^{\mathrm{book}}\) is the net delta of the book, then \(\Delta_t^{\mathrm{target}}=-\delta_t^{\mathrm{book}}\). The agent chooses the target stock holding
\[
    h_t = a_t \Delta_t^{\mathrm{target}} = -\,a_t\,\delta_t^{\mathrm{book}}.
\]
Thus, \(a_t=1\) corresponds to delta hedging, \(0\le a_t<1\) corresponds to under-hedging relative to delta neutrality, and \(a_t>1\) corresponds to over-hedging. The clipping range allows the policy to reduce the intensity of the hedge, partially reverse the hedge (for \(a_t<0\)), or take a more aggressive hedge when the learned risk objective favors doing so.

The reward is the one-step marked-to-market PnL of the hedged portfolio net of transaction costs. Let \(V_t^{\mathrm{book}}\) denote the signed value of the client option book from the perspective of the desk, and let \(S_t\) denote the underlying price. After choosing \(h_t\), the per-step reward combines the stock-hedge PnL, the option-book PnL and the proportional trading cost \(c\,|h_t-h_{t-1}|\,S_t\) with \(c=3\) bps:
\[
    r_t = h_t(S_{t+1}-S_t) + \big(V_{t+1}^{\mathrm{book}} - V_t^{\mathrm{book}}\big) - c\,|h_t-h_{t-1}|\,S_t.
\]
Both values cover the contracts held over the step, with expiring contracts at intrinsic; arrivals are booked afterwards at the mid-quote and create no PnL. This accounting-style reward assigns the consequence of each hedge decision to the next marked-to-market change on a daily basis, making the learning signal more immediate than waiting until option expiration.

Because the cost depends on the previous holding \(h_{t-1}\), which is not part of \(o_t\), the agent cannot exactly anticipate the cost of its own rebalancing.

For a policy \(\pi\), the return from time \(t\) is
\[
    G_t^\pi = \sum_{k=t}^{T-1}\gamma^{k-t}r_k.
\]
We optimize the risk functionals of the distribution of this return rather than its mean (Section~\ref{subsec:dtd3}).

\subsection{Distributional TD3}
\label{subsec:dtd3}

Our main learning algorithm is a distributional variant of TD3. The actor \(\pi_\phi:\mathcal{O}\to\mathcal{A}\), with parameters \(\phi\), maps the {observation} \(o_t\) defined above to a continuous hedge ratio. The replay buffer \(\mathcal{D}\) stores observation transitions \((o_t,a_t,r_t,o_{t+1})\). Instead of estimating only the conditional mean return, each critic estimates a quantile approximation to the full return distribution. We use two distributional critics, with parameters \(\theta_1,\theta_2\),
\[
    Z_{\theta_1}(o,a), \qquad Z_{\theta_2}(o,a),
\]
where each critic outputs \(N\) quantile values at fixed quantile levels \(\{\tau_n\}_{n=1}^N\) (with \(N=99\), \(\tau_n=n/100\)). The distributional representation allows VaR and CVaR objectives to be computed directly from the learned return distribution.

Given a transition \((o_t,a_t,r_t,o_{t+1})\), the target actor produces the next action with TD3 target policy smoothing: clipped Gaussian noise is added to the output of the target actor and the result is clipped back to the valid action range,
\[
\begin{aligned}
\tilde{a}_{t+1}
&=
\mathrm{clip}\!\left(
    \pi_{\bar{\phi}}(o_{t+1})
    +
    \mathrm{clip}\big(\epsilon,\,-c_{\mathrm{tps}},\,c_{\mathrm{tps}}\big),
    \;a_{\mathrm{low}},\;a_{\mathrm{high}}
\right), \\[0.4em]
\epsilon &\sim \mathcal{N}(0,\sigma_{\mathrm{tps}}^2).
\end{aligned}
\]
with \(\sigma_{\mathrm{tps}}=0.2\), \(c_{\mathrm{tps}}=0.5\), and \([a_{\mathrm{low}},a_{\mathrm{high}}]=[-0.2,2.0]\). The target critics then estimate the next-step return distribution. To reduce over-estimation, we construct a conservative distributional target using the element-wise minimum of the two target critics:
\[
    \bar{z}_n
    =
    \min\left\{
        \theta^{-}_{1,n}(o_{t+1},\tilde{a}_{t+1}),
        \theta^{-}_{2,n}(o_{t+1},\tilde{a}_{t+1})
    \right\},
    \qquad n=1,\ldots,N.
\]
The distributional Bellman target is the one-step target
\[
    y_n = r_t + \gamma \bar{z}_n,
\]
with the continuation set to zero at the horizon.

Both critics are trained to match the target quantile distribution using the quantile Huber loss. For the critic \(i\in\{1,2\}\), the loss is
\[
    \mathcal{L}_i(\theta_i)
    =
    \frac{1}{N^2}
    \sum_{m=1}^{N}\sum_{n=1}^{N}
    \rho^{\kappa}_{\tau_m}
    \left(
        y_n - \theta_{i,m}(o_t,a_t)
    \right),
\]
where \(\rho^{\kappa}_{\tau}\) is the standard quantile Huber loss at the quantile level \(\tau\) (Huber parameter \(\kappa=1\)). Taking the element-wise minimum of the two critics guards against overestimating the sparsely observed extreme quantiles.

The update schedule decouples the four sets of parameters. The two online critics are updated every step of the learner. The online actor is updated with a delayed schedule, once every \(d=2\) critic steps, and the target actor is soft-updated by Polyak averaging with coefficient \(\tau_{\mathrm{actor}}=0.002\) every 2 steps. Unlike standard TD3, we do not update the target critics whenever the delayed actor is updated. Instead, we apply soft updates to target critics every 100 steps using \(\tau_{\mathrm{critic}}=0.002\). This slower schedule keeps the estimated return distribution stable while the policy adjusts; we use a 100-step interval, slow enough to stabilize the distributional target, but fast enough to track the improving policy.

The actor objective is to choose hedge ratios that maximize a selected risk functional of the learned return distribution, evaluated on the element-wise minimum of the two critics:
\[
    J(\phi)
    =
    \mathbb{E}_{o\sim\mathcal{D}}
    \left[
        \rho\!\left(
            \min\big\{Z_{\theta_1}(o,\pi_\phi(o)),\,Z_{\theta_2}(o,\pi_\phi(o))\big\}
        \right)
    \right],
\]
where \(\mathcal{D}\) is the replay buffer, the minimum is taken element-wise across quantiles, and \(\rho\) is the selected risk measure. We evaluate policies trained with VaR\(_{95}\), CVaR\(_{95}\) and VaR\(_{99}\). For VaR\(_{95}\), \(\rho\) returns the \(5\%\) quantile of the predicted return distribution of the critic; for CVaR\(_{95}\), \(\rho\) aggregates the quantile values in the lower 5\% tail as an unnormalized sum, which is a positive multiple of the tail mean and therefore induces the same actor gradient direction. We do not train a separate CVaR\(_{99}\) agent: with 99 quantile levels the lower 1\% tail of the predicted distribution contains a single quantile, so the CVaR\(_{99}\) objective is a positive multiple of the VaR\(_{99}\) objective and therefore induces the same actor gradient direction. The training objective \(J\) averages the tail risk across the observations sampled from the replay buffer, so it is a surrogate for the CVaR at episode-level that we report.

\paragraph{Implementation.} The policy is a layer-normalized MLP (256-256-256) and each critic is a layer-normalized MLP (512-512-256) with a quantile head. All networks are trained with Adam with a learning rate of $10^{-4}$, batch size 256, a replay buffer of $10^6$ transitions, and constant Gaussian exploration noise ($\sigma=0.3$). Each agent is trained for 60,000 episodes and we evaluate the final policy without early stopping. Observation normalization uses fixed constants during training. We train the empirical CVaR$_{95}$ agent and the hybrid agent five times each with independent random seeds, and every other configuration three times.

\section{Empirical Evaluation}

\subsection{Evaluation Protocol and Measures}

\paragraph{Experimental setup.}
Agents are trained on data from 1996--2017 and evaluated out of sample from January 2018 to August 2023. The period includes the COVID-19 crash and the 2022 rate-hike selloff, along with other stable market regimes. All training runs employ the same arrival process, transaction cost specifications, reward function, episode length, and historical testing environment settings. Cross-objective studies utilize the risk metrics listed in Table~\ref{tab:robustness_all} to train agents. Each policy is evaluated against a delta-hedging benchmark in 40-day test windows. We also test the robustness and interpretability in addition to the performance of the policy in several ways. We compare it against a set of simple rules to see if the policy adds value, look at results across training environments and risk objectives to test consistency, and examine its state-dependent exposure to understand how it reacts in different situations.

\paragraph{Delta-hedging benchmark.}
Our primary benchmark is the delta hedging strategy. At the end of each day, the delta agent compensates for the net delta of the client book by buying or selling the underlying index, which means that it performs the action \(a_t = 1\) at each step. The evaluation of this benchmark follows the same process as RL agents, using the same episode sampling, meaning the same evaluation windows and client books, to ensure that all comparisons are made episode by episode.

\paragraph{Risk measures.}
For policy \(\pi\), the total PnL in episode \(i\) is
\[
    R_i^\pi = \sum_{t=0}^{T-1} r_{i,t}^{\pi}.
\]
We report the mean and standard deviation for episode PnL, together with VaR and CVaR as objective risk measures. In this section, we mainly compare CVaR\(_{95}\) between different agents, that is, the average PnL within the worst 5\% of episodes, since this is our main metric and the agent is optimized based on the CVaR\(_{95}\) training objective. To address test window overlap, we report confidence intervals from a bootstrap clustered by window start date and from a 63-trading-day calendar-block bootstrap.
\[
\begin{aligned}
    \mathrm{VaR}_{95}(R^\pi)
        &= q_{0.05}(R^\pi), \\
    \mathrm{CVaR}_{95}(R^\pi)
        &= \mathbb{E}\!\left[
            R^\pi \mid R^\pi \leq \mathrm{VaR}_{95}(R^\pi)
        \right].
\end{aligned}
\]

\paragraph{Net delta.}
To measure portfolio exposure after each hedging step, we use the net delta. The net delta measures how the hedged portfolio still co-moves with the underlying: it is the part of the client-book delta left unhedged by the RL policy,
\[
    \delta_t^{\mathrm{net}} = (1-a_t)\,\delta_t^{\mathrm{book}}.
\]

\paragraph{Market regimes.}
The evaluation episodes can be segmented into four quadrants based on the cumulative return of the realized SPX price path and its volatility. The cumulative return is
\[
    G_i^S = \frac{S_{i,T}-S_{i,0}}{S_{i,0}},
\]
and \(\sigma_i\) is the volatility of the daily underlying returns in the 40-day window. Episodes above the evaluation sample's 90th volatility percentile, \(\sigma_{p90} = 0.263\), are classified as high volatility; the sign of \(G_i^S\) separates rising and falling markets. This lets us group the episodes into four regimes: Q1, high-volatility rallies (e.g., the post-COVID recovery); Q2, high-volatility crashes (e.g., the March 2020 COVID-19 selloff); Q3, low-volatility slow declines (e.g., the 2022 rate-hike selloff); and Q4, low-volatility slow bull markets (e.g., the calmer remainder of the sample). We examine the policy's behavior in each regime.

\subsection{Performance Relative to Delta Hedging}
\label{subsec:perf}
The TD3 agent shows better performance compared to delta hedging in CVaR$_{95}$. Pooling five independent training runs, CVaR$_{95}$ increases from $-80.2$ to $-77.7$, giving a $2.5$-point improvement, which is equal to $3.1\%$ of delta-hedge tail loss (Table~\ref{tab:resid_td3_real}); every run improves. Improvements are primarily concentrated during downtrends, with tail losses decreasing by $8.9\%$ in Q2 and $4.4\%$ in Q3. In contrast, the strategy loses $7.4\%$ during stable uptrends (Q4) and $9.9\%$ during volatile uptrends (Q1). We call this protection cost the insurance premium of the strategy and measure it as the average PnL the strategy gives up relative to delta hedging: $0.65$ per episode, with an unchanged standard deviation, net of losses during uptrends and gains during downtrends.

\begin{table}[h]
\centering
\caption{Net delta and CVaR$_{95}$ by market regime (empirical TD3, five training runs pooled).}
\label{tab:resid_td3_real}
{\small
\begin{tabular}{@{}lrrrrr@{}}
\toprule
\textbf{Regime} & \textbf{Net~$\delta$} & \textbf{RL} & \textbf{Delta} & \textbf{Impr.} & \textbf{Impr.~\%} \\
\midrule
Q1 HV$\uparrow$   & $-3.4$ & $-134.6$ & $-122.5$ & $-12.1$ & $-9.9\%$ \\
Q2 HV$\downarrow$ & $-7.1$  & $-184.0$ & $-201.9$ & $+17.9$ & $+8.9\%$ \\
Q3 LV$\downarrow$ & $-7.5$ & $-64.6$ & $-67.6$ & $+3.0$  & $+4.4\%$ \\
Q4 LV$\uparrow$   & $-5.0$ & $-54.7$ & $-50.9$ & $-3.8$  & $-7.4\%$ \\
\midrule
\textbf{All}      & $-5.7$ & $-77.7$ & $-80.2$ & $+2.5$  & $+3.1\%$ \\
\bottomrule
\end{tabular}}
\tabnote{Net~$\delta$ is mean net delta (negative $=$ net short). HV/LV denote the high/low-volatility regimes, and the arrows the sign of the cumulative market return. Impr.\ is RL minus delta-hedge CVaR$_{95}$; Impr.~\% scales this difference by the delta-hedge tail loss.}
\end{table}

\paragraph{Turnover and transaction costs.}
During the evaluation, the RL strategy has approximately $3\%$ less trading volume than the delta hedging strategy ($1{,}468$ vs.\ $1{,}520$ shares). The average cumulative trading cost of the RL strategy is $4.43$, while that of the delta hedging strategy is $4.59$, a difference of $0.16$. Since these savings are negligible compared to the $2.5$-point improvement in CVaR$_{95}$, the lower turnover rate cannot explain the tail-risk outcome. This strategy does not use the trading cost rate as input, so we can recalculate its profit and loss using the same actions at different cost rates. For cost rates from $0$ to $20$ basis points, the strategy consistently maintains a CVaR$_{95}$ improvement of $1.9$ to $4.6$ points over the delta hedging strategy in multiple runs. Furthermore, this strategy outperforms the delta hedging strategy in both the worst $5\%$ episodes under the delta hedging strategy and the worst $5\%$ episodes under its own strategy.

A bootstrap clustered by the $1{,}386$ window start dates gives a $95\%$ confidence interval of $[+1.1,+3.8]$ points for the multi-run CVaR$_{95}$ improvement. Because all windows come from one realized path, a 63-day calendar-block bootstrap gives a wider interval, $[-2.1,+8.1]$. Under this more conservative interval, the improvement is not statistically distinguishable from zero, although all five training runs improve on this test history.

\subsection{Simple Hedge-Ratio Rules}
\label{sec:rules}
A strategy could reduce tail losses by holding a fixed short hedge. The learned policy does carry such a net-short tilt, whose state dependence we analyze in Section~\ref{subsec:tilt}. Therefore, we compare the learned strategy to six rules with parameters tuned to maximize CVaR$_{95}$ of delta-hedge episodes during the training period. These rules include a fixed hedge ratio $a$, a sign adjustment $k$ that under-hedges short books and over-hedges long books, different hedge ratios for short and long books, a fixed short offset of $m$ net-delta units, a no-trade band of width $b$, and a band around $0.95$ times delta. The RL agent outperforms all rules selected during the training period (Table~\ref{tab:rules}), while the best rule only recovers about half of the gain. This gain is not explained by the average short position alone, as the agent earns $1.2$ points more than a fixed short offset that matches its own average net delta (start-date-clustered interval $[+0.6,+1.9]$). The optimal fixed ratio ($a=0.97$) allows the purchase of a small crash payoff (Q2 $+4.9$) at virtually zero cost in a rising market, while the agent purchases a much larger payout (Q2 $+17.9$) and pays the premium described in Section~\ref{subsec:perf}.

\begin{table}[t]
\centering
\caption{Hedge-ratio rules chosen on 1996--2017 and evaluated on 2018--2023: change vs.\ delta hedging.}
\label{tab:rules}
{\small
\setlength{\tabcolsep}{2.4pt}
\begin{tabular}{@{}lrrrrrr@{}}
\toprule
\textbf{Rule (parameter)} & \textbf{CVaR$_{95}$} & \textbf{Q2} & \textbf{Q4} & \textbf{CVaR$_{99}$} & \textbf{Mean} & \textbf{Net~$\delta$} \\
\midrule
Constant ratio ($a=0.97$)        & $+1.1$ & $+4.9$ & $-0.0$ & $+2.8$ & $+0.21$ & $-0.8$ \\
Sign rule ($k=0.02$)             & $+0.0$ & $+2.1$ & $-1.5$ & $+0.9$ & $-0.49$ & $-2.7$ \\
Two-parameter ($0.96$, $1.00$)   & $+0.8$ & $+5.5$ & $-1.4$ & $+2.9$ & $-0.35$ & $-3.2$ \\
Short offset ($m=5$)             & $+1.1$ & $+8.7$ & $-2.7$ & $+4.5$ & $-0.81$ & $-5.0$ \\
No-trade band ($b=2$)            & $-0.0$ & $-0.1$ & $+0.0$ & $-0.0$ & $+0.01$ & $0.0$ \\
Band around $0.95\delta$ ($b=5$) & $+1.3$ & $+7.5$ & $-0.5$ & $+3.9$ & $+0.41$ & $-1.4$ \\
\midrule
\textbf{RL}                      & $\mathbf{+2.5}$ & $\mathbf{+17.9}$ & $-3.8$ & $\mathbf{+8.9}$ & $-0.65$ & $-5.7$ \\
\bottomrule
\end{tabular}}
\tabnote{Columns are changes vs.\ delta hedging in CVaR$_{95}$ (overall, Q2 and Q4), CVaR$_{99}$ and mean PnL; positive favors the rule. Same test windows and five RL runs as Table~\ref{tab:resid_td3_real}.}
\end{table}
\subsection{Training Environment Comparison}
Table~\ref{tab:training_learner_2x2} compares TD3 agents with a common CVaR$_{95}$ objective trained in empirical, GBM, oracle-$\sigma$ GBM, SABR, and hybrid environments, all evaluated in the same historical replay 2018--2023.

\begin{table}[h]
\centering
\caption{Training-environment comparison (TD3, CVaR$_{95}$, three training runs each).}
\label{tab:training_learner_2x2}
{\small
\setlength{\tabcolsep}{2.4pt}
\begin{tabular}{@{}lrrrrrrc@{}}
\toprule
& \multicolumn{5}{c}{\textbf{Net~$\delta$ by regime}} & \multicolumn{2}{c}{\textbf{CVaR$_{95}$ Impr.~\%}} \\
\cmidrule(lr){2-6} \cmidrule(l){7-8}
\textbf{Training} & \textbf{Q1} & \textbf{Q2} & \textbf{Q3} & \textbf{Q4} & \textbf{All} & \textbf{Mean} & \textbf{Range} \\
\midrule
Empirical & $-3.5$ & $-7.1$ & $-7.7$ & $-5.4$ & $\mathbf{-6.1}$ & $\mathbf{+3.3\%}$ & $[+2.9,+3.8]$ \\
GBM       & $+0.4$ & $+0.4$ & $+0.5$ & $+0.5$ & $+0.5$ & $-1.0\%$ & $[-1.5,-0.1]$ \\
GBM (oracle $\sigma$) & $+0.0$ & $+0.0$ & $+0.1$ & $+0.1$ & $+0.1$ & $+0.0\%$ & $[-1.2,+0.8]$ \\
SABR      & $-3.3$ & $-4.9$ & $-4.8$ & $-4.1$ & $-4.3$ & $+1.1\%$ & $[+0.3,+1.7]$ \\
Hybrid    & $-2.6$ & $-2.6$ & $-3.0$ & $-2.8$ & $-2.8$ & $+1.9\%$ & $[+1.5,+2.2]$ \\
\bottomrule
\end{tabular}}
\tabnote{Net~$\delta$: mean net delta by regime (negative $=$ net short). Impr.~\%: CVaR$_{95}$ improvement over delta hedging as a share of its tail loss, mean and range across runs, on the test windows of Table~\ref{tab:resid_td3_real} (the Empirical and Hybrid rows use three of their five runs).}
\end{table}

The table shows that, among all training environments, empirical training produces the greatest improvement, $+3.3\%$ against $-1.0\%$ for GBM training. This suggests that the source of the training data drives much of the improvement in tail risk.

Across the three training runs common to all environments, the empirical agent has a net delta of $-6.1$ and a tail-loss reduction of $3.3\%$ ($-5.7$ and $3.1\%$ over all five runs, Table~\ref{tab:resid_td3_real}), with a net negative delta in all regimes. In contrast, the GBM-trained TD3 strategy has a slightly positive net delta throughout, close to delta hedging, and performs about $1.0\%$ worse than delta hedging. We also train a GBM agent with the volatility realized during the evaluation period ($\sigma=0.21$), information that is not available in real time; because the arrival trigger is one daily standard deviation, it also moves from $1.20\%$ to $1.32\%$. Even with this look-ahead advantage, the strategy still keeps the net delta near zero in all cases, only matching the delta hedging strategy ($+0.0\%$), so the difference is not due to the volatility level alone. The hybrid agent, trained with the same historical paths and listed contracts but using Black--Scholes option valuation, underperforms the empirical agent. It improves by $1.9\%$, with a smaller and almost constant tilt (net delta of $-2.8$), and its CVaR$_{95}$ decreases by $0.8$ points over five training runs (start-date-clustered interval $[-1.6,-0.1]$; $[-3.8,+1.5]$ with 63-day calendar blocks). It still outperforms the delta hedging strategy, so the historical paths contribute most of the gain; valuing the book with market quotes also increases the gain in our runs, but not significantly under calendar blocks. For the agent trained with SABR, stochastic volatility leads to a smaller short tilt (net delta of $-4.3$) and a $1.1\%$ reduction in CVaR$_{95}$ tail loss, but its effect is not superior to that of a fixed short tilt of the same magnitude. The stochastic-volatility simulator we tested therefore recovers only part of the effect of historical paths.

\subsection{State-Dependent Net Exposure}
\label{subsec:tilt}
\begin{figure}[t]
\centering
\includegraphics[width=\linewidth]{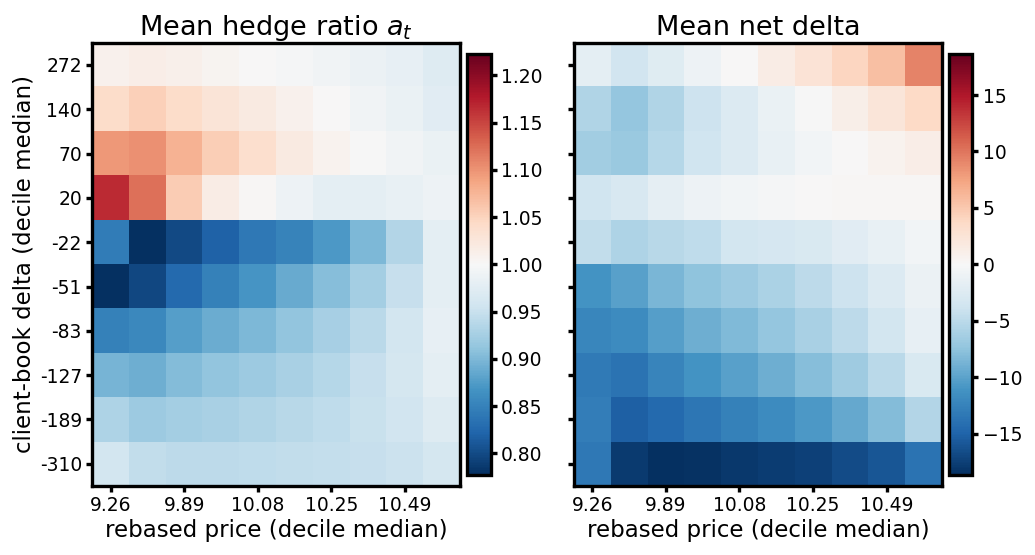}
\caption{State-dependent net exposure (empirical TD3, CVaR$_{95}$, five runs pooled): mean hedge ratio (left) and net delta (right) by client-book delta and price decile.}
\Description{Heatmaps showing the learned hedge ratio and net delta across client-book delta and price deciles.}
\label{fig:hedging_tilt}
\end{figure}

The empirically trained TD3 policy hedges most of the book's delta on average, but leaves a consistently negative net delta. Its average hedge ratio is approximately $0.95$, while its mean net delta is $-5.7$. Figure~\ref{fig:hedging_tilt} pools the five training runs.

This policy attains its negative net delta primarily via the short client books. When the client book has a negative delta, the policy under-hedges by buying less stock than a full delta hedge would, leaving the hedged portfolio net short on delta. Conversely, when the book is exposed to a positive delta, the policy flips its hedge action, hedging almost fully and over-hedging only small long books after the index has fallen. Figure~\ref{fig:hedging_tilt} shows the policy tilting the portfolio's net delta exposure toward a position that gains when the underlying falls. The pattern is strongest in the extreme negative delta bins and becomes more pronounced at lower prices after the index has fallen within the window. This behavior is difficult for a simple rule overlay to replicate across both rising and falling markets.

\subsection{Robustness Across Risk Objectives}
\label{sec:robustness}
We retrain empirical TD3 strategies under VaR$_{95}$ and VaR$_{99}$, in addition to CVaR$_{95}$, and then evaluate each strategy using its respective risk metric (see Table~\ref{tab:robustness_all}). Each objective is retrained in three independent runs (five for CVaR$_{95}$), so the improvements reported here are the averages of these runs; the row for CVaR$_{95}$ corresponds to Table~\ref{tab:resid_td3_real}. We observe that all three objectives show a negative net delta. For CVaR$_{95}$, the improvement remains in a positive range, about $2.0$--$3.1$ points, outperforming delta hedging in every run. VaR$_{95}$ only slightly improves its own metric, but increases CVaR$_{95}$ by $1.9$ points without additional cost in mean PnL. VaR$_{99}$ matches delta hedging on its own measure and is not better than short fixed offsets of the same scale.

\begin{table}[h]
\centering
\caption{Robustness across risk objectives (each policy evaluated on its own objective).}
\label{tab:robustness_all}
{\small
\begin{tabular}{@{}lrrrrc@{}}
\toprule
\textbf{Objective} & \textbf{Net~$\delta$} & \textbf{Impr.} & \textbf{Impr.~\%} & \textbf{Std.} & \textbf{Range} \\
\midrule
CVaR$_{95}$ & $-5.7$ & $+2.5$ & $+3.1\%$ & $0.4$ & $[+2.0,\,+3.1]$ \\
VaR$_{95}$  & $-2.5$  & $+0.5$ & $+1.0\%$ & $0.1$ & $[+0.4,\,+0.6]$ \\
VaR$_{99}$  & $-10.4$ & $+0.4$ & $+0.4\%$ & $1.5$ & $[-0.6,\,+2.0]$ \\
\bottomrule
\end{tabular}}
\tabnote{Impr.: RL minus delta hedging on each objective's own measure, mean over training runs; Impr.~\% scales it by the delta-hedge loss; Std.\ and Range: dispersion of the improvement across runs.}
\end{table}

\subsection{Stress-Period Case Studies}
\begin{figure}[h]
\centering
\includegraphics[width=\linewidth]{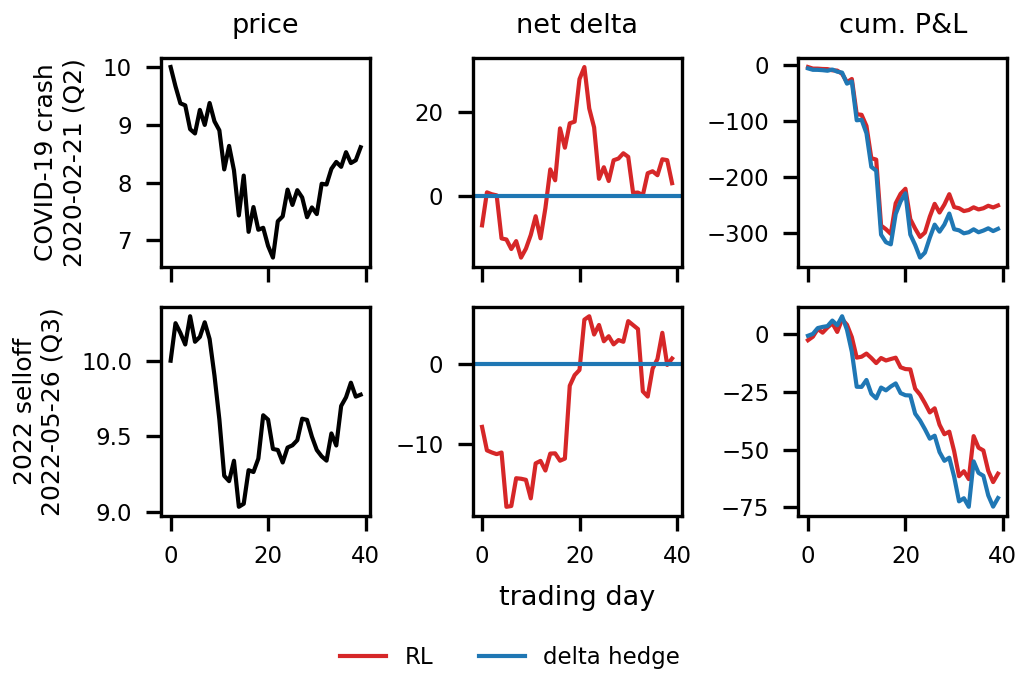}
\caption{Stress-period paths for empirical TD3 (CVaR$_{95}$): COVID-19 crash (Q2, top) and 2022 selloff (Q3, bottom).}
\Description{Price, net delta, and cumulative PnL for empirical TD3 and delta hedging during two held-out stress episodes.}
\label{fig:episode}
\end{figure}

Figure~\ref{fig:episode} traces the empirical TD3 policy (one training run) through the COVID-19 2020 crash and the slower 2022 rate-hike selloff, using the test window with the worst delta-hedge PnL in each.

During these two periods, the strategy ultimately ends up with a cumulative PnL higher than the delta hedging strategy. This exposure does not mean holding rigid short positions in stocks: in 2020, the net delta is negative during the first half of the decline and turns positive about a week before the market bottom as the client book turns long, but in 2022 it stays negative throughout the downturn. These examples illustrate the state-dependent behavior measured in the evaluation.

\section{Conclusion}
We explore whether training a risk-sensitive hedging agent based on real market data (rather than on a constant-volatility simulator) alters its learned strategy and out-of-sample tail risk. Two findings stand out. First, the distributional TD3 strategy trained in an empirical environment reduces tail loss by 3.1\% compared to delta hedging, outperforming all hedge-ratio rules selected on the training data. During periods of volatile market decline, the strategy reduces tail loss by 8.9\%; however, its performance is relatively poor during periods of market growth, resulting in a 7--10\% larger tail loss.

Second, empirical training gives the largest gain of the training environments we tested: the GBM-trained agent does not beat delta hedging, and the hybrid agent with Black--Scholes valuation and the SABR agent each recover only part of the gain. With a negative client-book delta, the policy under-hedges, most strongly after prices have fallen, which leaves a state-dependent negative net delta.

In summary, these results indicate that, for tail-risk objectives, the training environment can indeed influence what a hedging agent learns. However, this evidence is relatively limited, as it is based on only one historical test and its advantage is conditional: gains obtained in a declining market may be partially reversed in a rising market. For RL hedging agents, what the training market contains matters.

There are a few limitations that might affect these conclusions and raise questions for future research. First, empirical evidence is based only on a single out-of-sample period (2018--2023). Since the confidence intervals for the calendar blocks include zero, we need data from other markets and longer-term evaluations to draw broader conclusions. Next, we have not pinpointed which part of the conservative TD3 design is responsible for the tail-risk gain; controlled ablation experiments are left for future research. Finally, this environment is just a proof of concept. The natural next step is to include higher-order risk exposures (gamma and vega), multiple underlying assets, and richer trading costs to really show how far the empirical training advantages can carry into a full trading desk environment.

\bibliographystyle{ACM-Reference-Format}
\bibliography{rl_hedging_rev}

@article{black1973pricing,
  author  = {Black, Fischer and Scholes, Myron},
  title   = {The Pricing of Options and Corporate Liabilities},
  journal = {Journal of Political Economy},
  volume  = {81},
  number  = {3},
  pages   = {637--654},
  year    = {1973},
  doi     = {10.1086/260062}
}

@article{merton1973theory,
  author  = {Merton, Robert C.},
  title   = {Theory of Rational Option Pricing},
  journal = {The Bell Journal of Economics and Management Science},
  volume  = {4},
  number  = {1},
  pages   = {141--183},
  year    = {1973},
  doi     = {10.2307/3003143}
}

@article{leland1985option,
  author  = {Leland, Hayne E.},
  title   = {Option Pricing and Replication with Transactions Costs},
  journal = {The Journal of Finance},
  volume  = {40},
  number  = {5},
  pages   = {1283--1301},
  year    = {1985},
  doi     = {10.1111/j.1540-6261.1985.tb02383.x}
}

@inproceedings{lillicrap2016continuous,
  author    = {Lillicrap, Timothy P. and Hunt, Jonathan J. and Pritzel, Alexander and Heess, Nicolas and Erez, Tom and Tassa, Yuval and Silver, David and Wierstra, Daan},
  title     = {Continuous Control with Deep Reinforcement Learning},
  booktitle = {International Conference on Learning Representations},
  year      = {2016},
  eprint    = {1509.02971},
  archivePrefix = {arXiv},
  primaryClass  = {cs.LG}
}

@inproceedings{fujimoto2018addressing,
  author    = {Fujimoto, Scott and van Hoof, Herke and Meger, David},
  title     = {Addressing Function Approximation Error in Actor-Critic Methods},
  booktitle = {Proceedings of the 35th International Conference on Machine Learning},
  series    = {Proceedings of Machine Learning Research},
  volume    = {80},
  pages     = {1587--1596},
  year      = {2018}
}

@inproceedings{barth2018distributed,
  author    = {Barth-Maron, Gabriel and Hoffman, Matthew W. and Budden, David and Dabney, Will and Horgan, Dan and TB, Dhruva and Muldal, Alistair and Heess, Nicolas and Lillicrap, Timothy},
  title     = {Distributed Distributional Deterministic Policy Gradients},
  booktitle = {International Conference on Learning Representations},
  year      = {2018}
}

@inproceedings{bellemare2017distributional,
  author    = {Bellemare, Marc G. and Dabney, Will and Munos, R{\'e}mi},
  title     = {A Distributional Perspective on Reinforcement Learning},
  booktitle = {Proceedings of the 34th International Conference on Machine Learning},
  series    = {Proceedings of Machine Learning Research},
  volume    = {70},
  pages     = {449--458},
  year      = {2017}
}

@inproceedings{dabney2018distributional,
  author    = {Dabney, Will and Rowland, Mark and Bellemare, Marc G. and Munos, R{\'e}mi},
  title     = {Distributional Reinforcement Learning with Quantile Regression},
  booktitle = {Proceedings of the AAAI Conference on Artificial Intelligence},
  volume    = {32},
  number    = {1},
  pages     = {2892--2901},
  year      = {2018}
}

@article{buehler2019deep,
  author  = {Buehler, Hans and Gonon, Lukas and Teichmann, Josef and Wood, Ben},
  title   = {Deep Hedging},
  journal = {Quantitative Finance},
  volume  = {19},
  number  = {8},
  pages   = {1271--1291},
  year    = {2019},
  doi     = {10.1080/14697688.2019.1571683}
}

@article{kolm2019dynamic,
  author  = {Kolm, Petter N. and Ritter, Gordon},
  title   = {Dynamic Replication and Hedging: A Reinforcement Learning Approach},
  journal = {The Journal of Financial Data Science},
  volume  = {1},
  number  = {1},
  pages   = {159--171},
  year    = {2019},
  doi     = {10.3905/jfds.2019.1.1.159}
}

@article{cao2021deep,
  author  = {Cao, Jay and Chen, Jacky and Hull, John and Poulos, Zissis},
  title   = {Deep Hedging of Derivatives Using Reinforcement Learning},
  journal = {The Journal of Financial Data Science},
  volume  = {3},
  number  = {1},
  pages   = {10--27},
  year    = {2021},
  eprint  = {2103.16409},
  archivePrefix = {arXiv},
  primaryClass  = {q-fin.CP}
}

@article{cao2023gamma,
  author  = {Cao, Jay and Chen, Jacky and Farghadani, Soroush and Hull, John and Poulos, Zissis and Wang, Zeyu and Yuan, Jun},
  title   = {Gamma and Vega Hedging Using Deep Distributional Reinforcement Learning},
  journal = {Frontiers in Artificial Intelligence},
  volume  = {6},
  pages   = {1129370},
  year    = {2023},
  doi     = {10.3389/frai.2023.1129370}
}

@article{chen2024barrier,
  author  = {Chen, Jacky and Fu, Yu and Hull, John and Poulos, Zissis and Wang, Zeyu and Yuan, Jun},
  title   = {Hedging Barrier Options Using Reinforcement Learning},
  journal = {Journal of Investment Management},
  volume  = {22},
  number  = {4},
  pages   = {16--25},
  year    = {2024}
}

@article{mikkila2023empirical,
  author  = {Mikkil{\"a}, Oskari and Kanniainen, Juho},
  title   = {Empirical deep hedging},
  journal = {Quantitative Finance},
  volume  = {23},
  number  = {1},
  pages   = {111--122},
  year    = {2023},
  doi     = {10.1080/14697688.2022.2136037}
}

@article{ho1981optimal,
  author  = {Ho, Thomas and Stoll, Hans R.},
  title   = {Optimal Dealer Pricing under Transactions and Return Uncertainty},
  journal = {Journal of Financial Economics},
  volume  = {9},
  number  = {1},
  pages   = {47--73},
  year    = {1981}
}

@article{avellaneda2008high,
  author  = {Avellaneda, Marco and Stoikov, Sasha},
  title   = {High-Frequency Trading in a Limit Order Book},
  journal = {Quantitative Finance},
  volume  = {8},
  number  = {3},
  pages   = {217--224},
  year    = {2008}
}

@article{garleanu2009demand,
  author  = {G{\^a}rleanu, Nicolae and Pedersen, Lasse Heje and Poteshman, Allen M.},
  title   = {Demand-Based Option Pricing},
  journal = {The Review of Financial Studies},
  volume  = {22},
  number  = {10},
  pages   = {4259--4299},
  year    = {2009}
}

@inproceedings{spooner2018market,
  author    = {Spooner, Thomas and Fearnley, John and Savani, Rahul and Koukorinis, Andreas},
  title     = {Market Making via Reinforcement Learning},
  booktitle = {Proceedings of the 17th International Conference on Autonomous Agents and MultiAgent Systems (AAMAS)},
  pages     = {434--442},
  year      = {2018}
}

@article{hagan2002managing,
  author  = {Hagan, Patrick S. and Kumar, Deep and Lesniewski, Andrew S. and Woodward, Diana E.},
  title   = {Managing Smile Risk},
  journal = {Wilmott Magazine},
  volume  = {1},
  pages   = {84--108},
  year    = {2002}
}

@misc{zernikov2026deep,
  author        = {Zernikov, Kirill},
  title         = {What Does Deep Hedging Actually Learn? {D}elta Corrections, Regime Fragility, and Symbolic Distillation},
  year          = {2026},
  eprint        = {2605.21696},
  archivePrefix = {arXiv},
  primaryClass  = {q-fin.RM}
}

\end{document}